\documentclass[conference]{IEEEtran}
\IEEEoverridecommandlockouts
\usepackage{cite}
\usepackage{amsmath,amssymb,amsfonts}
\usepackage{algorithmic}
\usepackage{graphicx}
\usepackage{textcomp}
\usepackage{xcolor}
\def\BibTeX{{\rm B\kern-.05em{\sc i\kern-.025em b}\kern-.08em
    T\kern-.1667em\lower.7ex\hbox{E}\kern-.125emX}}
\begin{document}

\title{Nuclei panoptic segmentation and composition regression with multi-task deep neural networks}

\author{\IEEEauthorblockN{Satoshi Kondo}
\IEEEauthorblockA{Muroran Institute of Technology \\
Hokkaido, Japan \\
kondo@mmm.muroran-it.ac.jp}
\and
\IEEEauthorblockN{Satoshi Kasai}
\IEEEauthorblockA{Niigata University of Healthcare and Welfare \\
Niigata, Japan \\
satoshi-kasai@nuhw.ac.jp}
}

\maketitle

\begin{abstract}
Nuclear segmentation, classification and quantification within Haematoxylin \& Eosin stained histology images enables the extraction of interpretable cell-based features that can be used in downstream explainable models in computational pathology. The Colon Nuclei Identification and Counting (CoNIC) Challenge is held to help drive forward research and innovation for automatic nuclei recognition in computational pathology. This report describes our proposed method submitted to the CoNIC challenge. Our method employs a multi-task learning framework, which performs a panoptic segmentation task and a regression task. For the panoptic segmentation task, we use encoder-decoder type deep neural networks predicting a direction map in addition to a segmentation map in order to separate neighboring nuclei into different instances.

\end{abstract}


\section{Introduction}

Nuclear segmentation, classification and quantification within Haematoxylin \& Eosin stained histology images enables the extraction of interpretable cell-based features that can be used in downstream explainable models in computational pathology. The Colon Nuclei Identification and Counting (CoNIC) Challenge is held to help drive forward research and innovation for automatic nuclei recognition in computational pathology~\cite{conic}. The CoNIC challenge consists of two separate tasks. The first task (Task 1) is called “Nuclear segmentation and classification”. The purpose of this task is to segment nuclei within the tissue, while also classifying each nucleus into one of the following six categories: epithelial, lymphocyte, plasma, eosinophil, neutrophil or connective tissue. The second task (Task 2) is called “Task 2: Prediction of cellular composition”. The purpose of this task is to predict how many nuclei of each class are present in each input image. 

This report describes our method submitted to the CoNIC challenge. Our method employs a multi-task learning framework, which performs a panoptic segmentation task and  a regression task. For the panoptic segmentation task, we use an encoder-decoder type deep neural network. In the segmentation task, we introduce a direction map to separate neighboring nuclei into different instances. The predicted direction map is used in the postprocessing stage. For the regression task, we utilize the latent space representation obtained from the encoder of the segmentation network.

%%%
\section{Proposed Method}

\subsection{Network Architecture}

\begin{figure}[b]
\centerline{\includegraphics[scale=0.7]{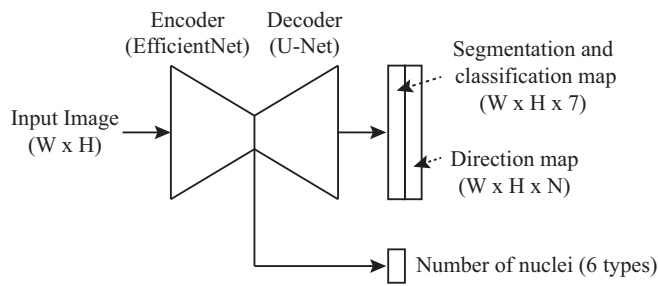}}
\caption{Network structure of our proposed method.}
\label{fig:network}
\end{figure}

Our method employs a multi-task learning framework, which performs a panoptic segmentation task (Task 1) and  a regression task (Task 2). Figure~\ref{fig:network} shows an overview of the network structure of our proposed method.

\begin{figure}[t]
\centerline{\includegraphics[scale=0.7]{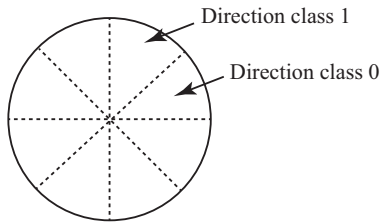}}
\caption{Example of direction map.}
\label{fig:direction}
\end{figure}

For the panoptic segmentation task, we use Eff-UNet~\cite{baheti2020eff} which combines the effectiveness of compound scaled EfficientNet~\cite{tan2019efficientnet} as the encoder for feature extraction with U-Net decoder~\cite{ronneberger2015u} for reconstructing the fine-grained segmentation map. Note that the last stage of EfficientNet is omitted. In Task 1, we have to segment nuclei within the tissue, while also classifying each nucleus into one of the following six categories: epithelial, lymphocyte, plasma, eosinophil, neutrophil or connective tissue. We use the Eff-UNet to predict two maps. The first map is a segmentation and classification map which has seven channels. Each channel has the same spatial size with input images. One of seven channels represents the background and each of the rest channels represents segmentation of each nucleus. The second map is a direction map. The encoding method to make the direction map was proposed by Uhrig et al.~\cite{uhrig2016pixel}.  For each pixel in a foreground object, its relative angle towards the centroid of the object is calculated, and quantized into $N$ different classes. An example of the direction map is shown in Fig.~\ref{fig:direction}. The direction map has $N$ channels and the background channel in the segmentation and classification map is used as the background for the direction map. The number of output channels of Eff-UNet is $7 + N$ in total.

For the regression task (Task 2), we utilize the latent space representation (feature vectors) obtained from the encoder of the segmentation and classification network. The feature vectors are fed into a network, which is the same as the last stage in EfficientNet, and then fed into a single linear layer. The number of output channels is six and each channel predicts the number of nuclei for each category.

\subsection{Training Procedure}

As the encoder of Eff-UNet, we use EfficientNet-B7 and it is pretrained with the ImageNet dataset~\cite{russakovsky2015imagenet}. The number of directions for the direction map, $N$, is 4. 

As for the preprocessing, the images are augmented. We use shift (maximum shift size is 10 \% of the image size), scaling (0.9 – 1.1 times), rotation (-5 – +5 degrees), color jitter (0.8 – 1.2 times for brightness, contrast, saturation and hue) and Gaussian blur (the max value of the sigma is 1.0) for the augmentation. The images are then normalized.

The loss function consists of four terms. The first term is the cross entropy loss for each pixel in the segmentation and classification map. The second term is the Dice loss. The background pixels are excluded in the Dice loss. The third term is the cross entropy loss for each pixel in the direction map. The background pixels are also excluded in this loss. The final loss is the L2 loss for the regression outputs. The weights for four terms are 1.0, 4.0, 2.0 and 0.005, respectively.

The datasets are constructed as follows. We use 80 \% of the images for training and 20 \% of the images for validation. The number of images in training and validation datasets are 3,983 and 998, respectively.

The optimizer is Adam~\cite{dp2015adam} and the learning rate changes with cosine annealing. The initial learning rate is 1e-3. The number of epoch is 30. The model with the highest value in Dice coefficients for the validation dataset is selected as the final model.

\subsection{Post Processing}

In the inference stage, the segmentation and classification map, the direction map and the number of nuclei for each category are obtained for each input image. 

From two maps, we reconstruct instance segmentation results and nuclei classification results. The instance segmentation results show nuclei area with the indices. The nuclei classification results show nuclei area with the class labels. At the first step, connected area having the direction class ID 0 are extracted. When $N$ is 4, we can extract right upper area of each nucleus. Hence we identify each nucleus and give an index to each nucleus. The next step is extraction of connected area having the direction class ID 1. The area with ID 0 and the area with ID 1 should be adjacent, as shown in Fig.~\ref{fig:direction}. We check if the area with ID 1 has adjacent area which ID is 0. When it is found, these areas are merged as the same nuculei. Otherwise, a new index is assigned to the area with ID 1. The same procedure is repeated until the direction class ID $N-1$ and we can obtain the instance segmentation results. Then we assign a nucleus class ID to each instance in the instance segmentation results by referring to the classification map. And we can obtain the nuclei classification results.

The number of nuclei for each category obtained by the network is a real value. If the value is negative, it is thresholded to 0. Otherwise, it is rounded to a nearest integer value. 

%%%
\section{Experimental Results}

The organizers of the CoNIC challenge prepare the preliminary test site to evaluate the performance for each task. As for the evaluation metrics for Task 1, multi-class panoptic quality ($mPQ$) is used to determine the performance of nuclear instance segmentation and classification.  $PQ$ is calculated with Detection Quality ($DQ$), which is calculated with numbers of TP, FP and FN, and Segmentation Quality ($SQ$), which is calculated with IoU and number of TP. For Task 2, multi-class coefficient of determination $R_{t}^{2}$ to determine the correlation between the predicted and true counts is used. For more details, please refer to the challenge site~\cite{conic}.

The results of the preliminary test of our proposed method were that $mPQ$ was 0.309 and $R_{t}^{2}$ was 0.344.

%\bibtyle{IEEEtran}
\bibliographystyle{IEEEtran}
\bibliography{references}

\end{document}